# Focusing by Plano-Concave lens using Negative Refraction

P. Vodo, P. V. Parimi[1], W. T. Lu, and S. Sridhar[2]
Electronic Materials Research Institute (EMRI) and
Physics Department, Northeastern University, 360 Huntington Avenue, Boston, Massachusetts 02115

We demonstrate focusing of a plane microwave by a plano-concave lens fabricated from a photonic crystal (PhC) having negative refractive index and left-handed electromagnetic properties. An inverse experiment, in which a plane wave is produced from a source placed at the focal point of the lens is also reported. A frequency dependent negative refractive index, $n(\omega)$ is obtained for the lens from the experimental data which matches well with that determined from band structure calculations.

Negative refraction in left-handed materials (LHM)[1,2,3,4,5,6] has triggered intense interest in designing novel microwave and optical elements, a flat lens being one of them[7]. In the case of the flat lens the waves entering from the source refract negatively on both interfaces and meet constructively on the far side of it. Thus a flat lens applies phase correction to the propagating waves similar to a conventional lens made of a naturally available material and having a positive index of refraction. However, it operates only when the source is close to the lens[7,8]. For a majority of applications of lenses in optics, astronomical telescopes, commercial and defense microwave communications, far field imaging is required. Negative refraction allows focusing of a far field radiation by concave rather than convex surfaces[9], with the advantage of reduced aberration for the same radius of curvature and changes many commonly accepted aspects of conventional optical systems. Of the two classes of LHM currently being investigated, focusing using a plano-concave lens made of a lefthanded metamaterial (MM) fabricated by interleaving arrays of wire strips and split ring resonators was demonstrated experimentally by Parazzoli et al[10].

In this letter we demonstrate that a real image of a far field radiation can be produced using a lefthanded PhC lens. We also report an inverse experiment in which the lens produces plane waves from a point source placed at the focal length. The frequency dependent refractive index, $n(\omega)$, determined from the experimental data is in complete agreement with that predicted by the theory using band structure calculations. The results confirm that far field focusing is realizable and open the door for several applications of the LHM in the far field region.

Microwave focusing measurement are carried out using three plano-concave lenses made of a dielectric PhC. The radii of curvature of the lenses are 13.5, 17.5 and 22 cm. The 2D PhC consists of a periodic array of alumina rods in air, arranged on a square lattice and having dielectric constant, $\varepsilon = 8.9$. The ratio of the radius of the alumina rods to the lattice spacing is $r/a = 0.175$. Microwave measurements are carried out in a parallel plate waveguide. An X-band waveguide kept at a distance of 150 cm from the lens' flat surface acts as a microwave source. The emitted wave travels through the parallel plate waveguide and the eventual plane wave is made to incident on the flat surface of the lens.

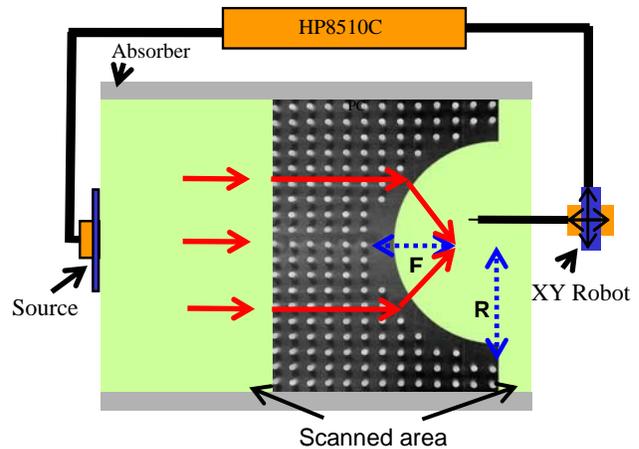

**Fig. 1** Schematic diagram of the microwave focusing experimental set up.

The propagation of the wave inside the PhC is along $\Gamma X$ direction of the Brillouin zone. Field maps of the incoming plane wave and the emerging radiation, on the far side, are captured using a monopole sensor on a ground plane. The sensor is hooked up to an automated X-Y translational stage which scans for the electric field component of the microwaves in the region of interest as shown in Fig. 1. An HP-8510C Network analyzer is used for measuring the transmission characteristics.

---

[1] Pa.parimi@neu.edu, [2] s.sridhar@neu.edu

Fig. 2 shows a sharp focus point achieved at 10.1cm using the plano-concave PhC lens of radius of curvature 13.5 cm. From left to right, the incoming plane wave, a real picture of the PhC lens and the emerging mapped field are shown. A clear focusing point is observed in the frequency range 9.265-9.490 GHz. Note that the direction of the energy flow changes only at the second interface of the plano-concave lens.

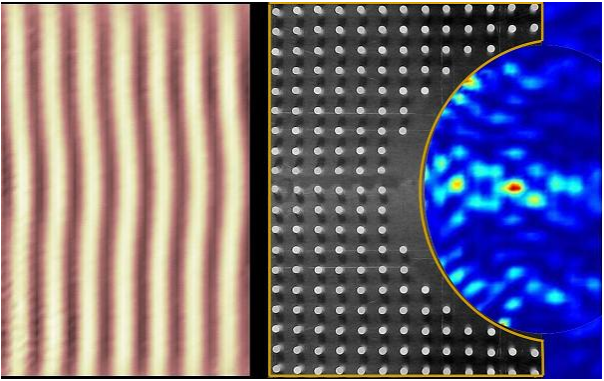

**Fig. 2** Focusing by a plano-concave PhC lens having radius of curvature $13.5$ cm. The focus point observed at $9.31$ Ghz is $10.1$ cm from the concave lens surface. A photograph of the PhC is superimposed on two Matlab surface plots to obtain the final figure. Dark strip in the center is a schematic representation of the area between the lens and the incoming wave. On the left side, field map of the incoming plane wave is shown (real part of transmission coefficient) and on the right side, intensity of the focus point. Scale: on the left, from $-0.025$ to $0.025$, on the right side from $0$ to $1.6 \times 10^{-3}$. Dimensions of the figure are $49 \times 34$ cm2. PhC lattice spacing is $1.8$ cm and the packing density of the square lattice is determined from the ratio r/a = $0.175$.

To validate that focusing is due to negative refraction an inverse experiment is carried out, in which a point source is kept at the observed focal point of the lens.

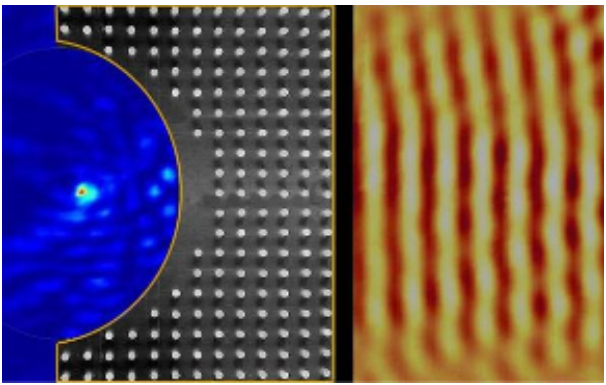

**Fig. 3** Field maps of the incident source and the emerging plane wave. Scale: on the left side intensity varies from $0.005$ to $0.055$, on the right side the real part of $S21$ from $-0.025$ to $0.015$. Dimensions of the figure are $49 \times 34$ cm2.

As can be seen from Fig. 3, a circular wave front from the point source after passing through the lens emerges as a plane wave. These two remarkable results validate the behavior of the lefthanded plano-concave lens. A similar lens made of a naturally available material, on the contrary would not be able to produce a real focus point as the present PhC lens does, but rather produce a diverging beam.

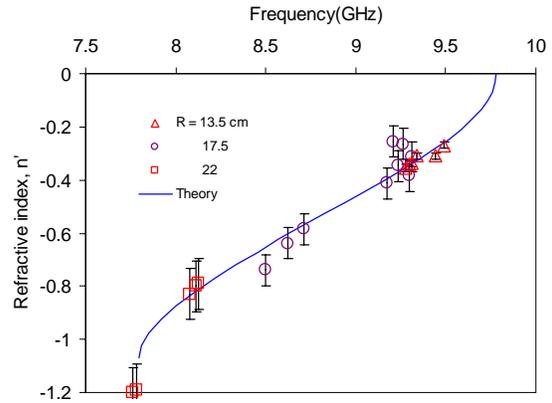

**Fig. 4** $n(\omega)$ determined from the focusing experiments using PhC lens. Note that the theoretical prediction (solid line) matches excellently with the experimental results.

The refractive index of the lens is determined using the lens equation $n = 1 - R/f$, where R is the radius of curvature and f is the focal length. Using this description we get $n = -0.4$ at 9.25 GHz. Note that real focusing by a plano-convex lens is achieved with $n > 1$, $R < 0$, while for the plano-concave lens with $R > 0$, $n < 1$. In Fig. 4 experimentally determined $n(\omega)$ for all three lenses is shown. It can be seen that for R=13.5 cm a sharp focus is achieved in the frequency range 9.25GHz to 9.5 GHz, for $R = 17.5$ cm in the range 8.5 GHz to 9.2 GHz and for $R = 22$ cm, 7.8 GHz to 8.1 GHz.

The nature of the lefthanded electromagnetism and focusing can be understood from the dispersion characteristics of the PhC. Fig. 5 shows band structure calculated for the PhC using a plane wave expansion method. From the band structure it can be deduced that in the second band for propagation along $\Gamma X$ direction, the wave vector, $k$ is in opposite direction to group velocity, $v_g$ : $v_g \cdot k < 0$ [11,5], resulting in negative refraction in the second band and correspondingly negative refractive indices. From the band structure theoretical refractive index $n = ck/\omega$ is determined. Solid line in Fig. 4 shows $n(\omega)$ obtained at various frequencies. As the frequency increases from 7.8-9.8 GHz the theoretical n increases from -1.1 to 0. Note that both the experimentally obtained refractive indices and the theoretically calculated values are in complete agreement.

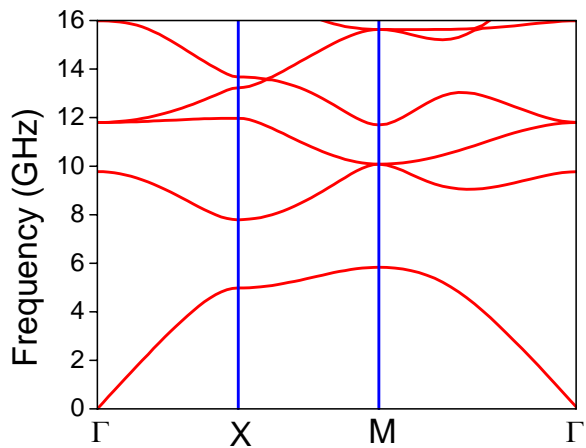

**Fig. 5** Band structure of the PhC calculated from plane wave expansion method.

In Fig. 4 it can be observed that $n(\omega)$ determined for each radius of curvature fall in certain frequency windows, and these windows move downwards as the radius of curvature is increased. The inhomogeneity of the PhC lens leads to corrugation on the concave surface. Windowing of refractive indices for each radius of curvature could be due to corrugation, as the conservation of wave vector on the refracting surface is considerably effected from such corrugation leading to no focus point. However, in the regions where focusing is achieved, corrugation has minimal effect on the sharpness of the image.

The present lens has several advantages when compared to the one with positive index. Lenses with reduced geometric aberrations produce sharper image with enhanced resolution and find numerous applications. For any value of n < 0, the radius of curvature of the lefthanded PhC lens is always larger than that of its counter part, a positive index plano-convex lens. Larger radius of curvature gives the advantage of reduced aberration in the image formed. Secondly, a PhC lens having the same focal length as that of a conventional lens weighs far less, and is suitable for space applications. The tailor made refractive index achievable in PhC materials[12] allows further control on the focal length and thereby helps reduce the length of the optical systems.

Bandwidth for obtaining a sharp focus point is a crucial parameter that decides the eventual applications of the lefthanded lenses. The present PhC lens reveals a wide bandwidth of 2 GHz, which is 22.7% at the current operating frequencies. In comparison to a plano-concave lens made of the MM, the PhC bandwidth is much larger. Due to the resonant nature of the MM the bandwidth is usually restricted to a narrow region and the dispersion is stronger[13]. The weaker dispersion in the PhC makes it a better candidate for focusing a pulse or broadband radiation.

In conclusion the feasibility of designing a broadband lefthanded lens is experimentally demonstrated. Focusing of plane waves by the plano-concave PhC lens is achieved for three different radii of curvature. The focal length follows the standard laws of geometrical optics applied with negative refraction. Further, the measured values of refractive indices of the lens are in complete agreement with those determined from band structure calculations.

This work was supported by the Air Force Research Laboratories, Hanscom AFB under contract #F33615-01-1-1007, and the National Science Foundation.